\documentclass[prb, twocolumn, superscriptaddress, showpacs, english]{revtex4-1}
\usepackage[T1]{fontenc}
\usepackage[latin9]{inputenc}
\usepackage{float}
\usepackage{amssymb}
\usepackage{amsmath}
\usepackage{graphicx}
\usepackage{babel}
\usepackage{color}
\usepackage{comment}
\usepackage[colorlinks,citecolor=blue]{hyperref}
\allowdisplaybreaks[1]
\def\cp#1{\mathbf{#1}}
\usepackage{epstopdf}

\begin{document}

\title{Gaussian variational method to Fermi Hubbard model in one and two dimensions}
	
\author{Yue-Ran Shi}
\affiliation{Department of Physics and Beijing Key Laboratory of Opto-electronic Functional Materials and Micro-nano Devices, Renmin University of China, Beijing 100872, China}
\affiliation{Key Laboratory of Quantum State Construction and Manipulation (Ministry of Education), Renmin University of China, Beijing 100872,China}

\author{Yuan-Yao He}
\affiliation{Institute of Modern Physics, Northwest University, Xi'an 710127, China}
\affiliation{Shaanxi Key Laboratory for Theoretical Physics Frontiers, Xi'an 710127, China}

\author{Ruijin Liu}
\email{rjliu@ustb.edu.cn}
\affiliation{Institute of Theoretical Physics, University of Science and Technology Beijing, Beijing 100083, China}

\author{Wei Zhang}
\email{wzhangl@ruc.edu.cn}
\affiliation{Department of Physics and Beijing Key Laboratory of Opto-electronic Functional Materials and Micro-nano Devices, Renmin University of China, Beijing 100872, China}
\affiliation{Key Laboratory of Quantum State Construction and Manipulation (Ministry of Education), Renmin University of China, Beijing 100872,China}
\affiliation{Beijing Academy of Quantum Information Sciences, Beijing 100193, China}

\begin{abstract}
The study of ground-state properties of the Fermi-Hubbard model is a long-lasting task in the research of strongly correlated systems, and is crucial for the understanding of notable quantum phenomena including superconductivity and magnetism. Owing to the exponentially growing complexity of the system, a quantitative analysis usually demands high computational cost and is restricted to small samples, especially in two or higher dimensions. Here, we introduce a variational method in the frame of fermionic Gaussian states, and obtain the ground states of one- and two-dimensional attractive Hubbard models via imaginary-time evolution. We calculate the total energy and benchmark the results in a wide range of interaction strength and filling factor with those obtained via exact two-body results, the density matrix renormalization group based on matrix product states (MPS), and projector Quantum Monte Carlo (QMC) method. For both 1D and 2D cases, the Gaussian variational method presents accurate results for total energy with a maximum systematic error $\sim 4\%$ in the intermediate interaction region. The accuracy of these results has negligible dependence on the system size. We further calculate the double occupancy and find excellent agreement with MPS and QMC, as well as the experimental results of cold quantum gases in optical lattices. The results suggest that the Gaussian pairing state is a good approximation to the ground states of attractive Hubbard model, in particular in the strong and weak coupling limits. Moreover, we generalize the method to the attractive Hubbard model with a finite spin-polarization, which can be mapped to the repulsive interaction case via particle-hole transformation, and obtain accurate results of ground state energy and double occupancy. Our work demonstrates the ability of the Gaussian variational method to extract ground state properties of strongly correlated many-body systems with negligible computational cost, especially of large size and in higher dimensions.
\end{abstract}

\maketitle


\section{Introduction}
\label{sec:intro}

Fermi-Hubbard model is one of the most simple yet important systems of spinful fermions. Since its first proposal by John Hubbard in 1963~\cite{Hubbard1963}, the model has been intensively studied by many analytical and numerical methods, and is suggested to be closely related to many notable quantum many-body phenomena including superconductivity and magnetism~\cite{Mott1968,Tasaki1994,Tasaki1995,Imada1998}. Unfortunately, analytical solutions of this model are currently absent for the physically most relevant cases in two and higher dimensions. Alternatively, it is well known that the attractive Hubbard model can be solved in a numerically exact manner at zero temperature by the density matrix renormalization group based on matrix product states (MPS) in one dimension and projector Quantum Monte Carlo (QMC) method in any dimension with arbitrary filling.  Nevertheless, the complexity of these numerical techniques makes it difficult to see clearly the underlying physics. Especially, the rather expensive numerical simulations present little useful information about the superconducting ground-state wavefunction except a specific superposition of slater determinants whose number scales exponentially with the system size. The applicability of numerical method is more limited for the repulsive Hubbard model, which is of great interest as a potential candidate to host high-$T_c$ superconducting state. Thus, revealing the many-body structure of the ground state for the attractive and repulsive Hubbard models still remains an open and difficult problem, despite the availability of some precise numerical results.

As one of the wavefunction methods in physics, the variational method has been considered particularly valuable in many cases since it allows to unveil fundamental physical mechanisms with relatively simple wavefunctions. On the other hand, the Gaussian states~\cite{Weedbrook2012} constitute one of the most successful families of variational wavefunctions. They form the basis of the Bardeen-Cooper-Schrieffer (BCS) theory of superconductivity and have been applied not only to describe the ground state but also to understand the nature of phase transitions into broken symmetry phases as well as the non-equilibrium dynamics of the order parameter~\cite{Barankov2006,Yuzbashyan2006}. One significant advantage of the Gaussian states is that the expectation values of physical observables can be cheaply and efficiently computed as they obey Wick's theorem~\cite{Wick1950}, which allows one to re-express expectation values of an arbitrary product of operators in terms of product of pairs. Thus, the Gaussian variational method, combining the variational method with Gaussian states, can be applied to crack various correlated systems with negligible computational effort comparing to the many-body numerical techniques. Further improvement of the method via appropriate canonical transformations is also available. Moreover, by extending Gaussian states to generalized non-Gaussian forms~\cite{Tao2018,Liu2020,Wang2020}, richer and more accurate variational solutions can be obtained, even for Bose-Fermi mixed systems.

In this work, we implement Gaussian variational method to study the ground state of attractive and repulsive Fermi-Hubbard models with representative fermion fillings in one and two dimensions.
For the attractive case, we obtain the ground state via imaginary-time evolution starting from a BCS-like pairing state, and compute the total energy and double occupancy of the model. Via the benchmarks with MPS and QMC methods in one and two dimensions respectively, we find that the Gaussian variational method presents highly accurate results of total energy with a maximum systematic error $\sim 4\%$ in the intermediate interaction region for both cases. Besides, our results of double occupancy show good agreement with the experimental data of cold quantum gases in optical lattices~\cite{Boll2016,Kohl2016,Hartke2022}. The results suggest that the Gaussian pairing states can capture the essential information of the ground state of attractive Hubbard model.

Further, we apply the method to attractive systems with finite spin polarization, which can be mapped to a repulsive Hubbard model via particle-hole transformation. By using a random initial state for variation, we obtain accurate results of ground state energy and double occupancy for 1D and 2D systems. The results demonstrate the ability of the Gaussian variational method to study large-size repulsive Hubbard model in higher dimensions, without assuming any specific form of Gaussian correlation {\it a priori}, and suggests the great potential to extend this low-cost method to investigate more complicated systems, such as interacting-background Fermi polarons, and spin- or mass-imbalanced Fermi-Hubbard models.

The remainder of the paper is organized as follows. In Sec.~\ref{sec:Hamiltonian} we formulate the Gaussian variational method for the attractive Fermi-Hubbard model with equal spin population. In Sec.~\ref{sec:res}, we calculate the two-body binding energy first, then we determine the ground state energy and double occupancy for different dimensions and system sizes. In Sec.~\ref{sec:repul}, we generalize the method to an attractive Hubbard model with a finite spin polarization, which can be exactly mapped to a repulsive Hubbard model via the particle-hole transformation. All results presented in Secs.~\ref{sec:res} and \ref{sec:repul} show excellent agreement with those obtained via other numerical methods and/or cold atom experiments. In Sec.~\ref{sec:con} we summarize the main findings and discuss promising directions for future studies.

\section{Hamiltonian and Gaussian variational method}
\label{sec:Hamiltonian}
	
We consider the attractive Hubbard model of spin-$1/2$ fermions in a $D$-dimensional square lattice with the following Hamiltonian
\begin{eqnarray}
\hat{H}&=&-t\sum_{\langle \mathbf{ij}\rangle\sigma}c_{\mathbf{i}\sigma}^{\dag} c_{\mathbf{j}\sigma} + U \sum_{\mathbf{i}}\hat{n}_{\mathbf{i}\uparrow}\hat{n}_{\mathbf{i}\downarrow}
-\mu \sum_{\mathbf{i}\sigma} c_{\mathbf{i}\sigma}^{\dag}c_{\mathbf{i}\sigma},
\label{eqn:H}
\end{eqnarray}
where $\sigma=\uparrow,\downarrow$ is the spin index and $\hat{n}_{\bf i\sigma}$ denotes the density operator on the $\mathbf{i}$-th lattice site. The model parameters include the nearest-neighbor hopping $t$, on-site Coulomb interaction $U$, and the chemical potential $\mu$, which is varied to control the fermion filling. For this moment, we focus on the attractively interacting $(U<0)$ and spin-balanced case with fermion number $N_\uparrow=N_\downarrow=N$, and define the fermion filling factor $\rho \equiv N/V$ with $V = L^D$ the total number of sites (with half-filling corresponding to $\rho=1/2$). The restriction of spin population will be released in Sec.~\ref{sec:repul}. We take $t$ as the energy unit and lattice spacing as the length unit throughout the discussions below.

We first rewrite the Hamiltonian in momentum space
\begin{eqnarray}
\hat{H} = \sum_{\mathbf{k}\sigma} (\varepsilon_\mathbf{k} - \mu) c_{\mathbf{k}\sigma}^{\dag} c_{\mathbf{k}\sigma}
+\frac{U}{V}\sum_{\bf{kk'q}}c_{\bf{k}\uparrow}^{\dag}c_{\bf{q-k}\downarrow}^{\dag}
c_{\bf{q-k'}\downarrow}c_{\bf{k'}\uparrow} ,
\label{eqn:Hk}
\end{eqnarray}
where $c_{\mathbf{k}\sigma}=\frac{1}{\sqrt{V}}\sum\limits_{\mathbf{i}} e^{-i\mathbf{k}\cdot\mathbf{i}}c_{\mathbf{i}\sigma}$ and $c_{\mathbf{k}\sigma}^\dagger=\frac{1}{\sqrt{V}}\sum\limits_{\mathbf{i}} e^{i\mathbf{k}\cdot\mathbf{i}}c_{\mathbf{i}\sigma}^\dagger$ are the fermionic operators in momentum space, and the kinetic energy dispersion reads $\varepsilon_\mathbf{k}=-2\cos k$ and $\varepsilon_\mathbf{k}=-2(\cos k_x + \cos k_y)$ for 1D and 2D lattices, respectively.
	
We then apply the fermionic Gaussian state to approximate the ground state of the model Hamiltonian Eq.(\ref{eqn:Hk}). The fermionic Gaussian state is expressed by
\begin{eqnarray}
|\Psi_{\rm GS}\rangle=\hat{U}_{\rm GS}|0\rangle,
\label{GS_state}
\end{eqnarray}
where $\hat{U}_{\rm GS}=e^{i\frac{1}{4}\hat{\mathbf{A}}^{T}\boldsymbol{\xi} \hat{\mathbf{A}}}$ is the Gaussian unitary operator, and the operator vector $\hat{\mathbf{A}}=(a^{\uparrow}_{1,\mathbf{k}_1},\ldots,a^{\uparrow}_{1,\mathbf{k}_V},$ $a^{\downarrow}_{1,\mathbf{k}_1},\ldots,a^{\downarrow}_{1,\mathbf{k}_V},$ $a^{\uparrow}_{2,\mathbf{k}_1},\ldots,a^{\uparrow}_{2,\mathbf{k}_V},$ $a^{\downarrow}_{2,\mathbf{k}_1},\ldots,a^{\downarrow}_{2,\mathbf{k}_V})^{T}$ is defined via the Majorana operators $a^{\sigma}_{1,\mathbf{k}_j}=c_{\mathbf{k}_j,\sigma}^{\dag}+c_{\mathbf{k}_j, \sigma}$ and $a^{\sigma}_{2,\mathbf{k}_j}=i(c_{\mathbf{k}_j,\sigma}^{\dag}-c_{\mathbf{k}_j, \sigma})$, which satisfy the anti-commutation relation $\{a^{\sigma}_{\alpha,\mathbf{k}}, a^{\sigma}_{\beta,\mathbf{k}'} \} = 2 \delta_{\alpha\beta} \delta_{\mathbf{k,k'}}$. The variational parameter $\boldsymbol{\xi}$ is an antisymmetric Hermitian matrix. To eliminate the gauge degree of freedom in $\boldsymbol{\xi}$, it is convenient to introduce a covariant matrix~\cite{Tao2018}
\begin{eqnarray}
\gamma_{i,j}=\frac{i}{2} \langle \Psi_{\rm GS}\ | [\hat{A}_i, \hat{A}_j]|\Psi_{\rm GS}\rangle ,
\label{eqn:covariant}
\end{eqnarray}
where $\hat{A}_i$ is the $i$-th element of $\hat{\mathbf{A}}$.
The fermionic Gaussian state can be completely characterized by the corresponding covariant matrix $\boldsymbol{\gamma}$, which is related to $\boldsymbol{\xi}$ as
\begin{eqnarray}
\boldsymbol{\gamma}=\mathbf{P}\boldsymbol{\Omega}\mathbf{P}^T,
\end{eqnarray}
with $\mathbf{P}=e^{i\boldsymbol{\xi}}$ and the symplectic matrix $\boldsymbol{\Omega}$ as
\begin{eqnarray}
\boldsymbol{\Omega}=\left(
  \begin{array}{cc}
    \mathbf{0} & -\mathbf{1}_{2V} \\
    \mathbf{1}_{2V} & \mathbf{0} \\
  \end{array}
\right).
\label{fermi_sigma}
\end{eqnarray}

In the spirit of variational method, the ground state of a Hamiltonian $\hat{H}$ can be obtained via an imaginary-time evolution of
\begin{eqnarray}
|\Psi(\tau)\rangle&=&\frac{e^{-\hat{H}\tau}|\Psi(0)\rangle}{\sqrt{\langle\Psi(0)|e^{-2\hat{H}\tau}|\Psi(0)\rangle}}
\label{eqn:ima1}
\end{eqnarray}
to the asymptotic limit $\tau\rightarrow\infty$, provided that the initial trial state $|\Psi(0)\rangle$ has a nonzero overlap with the ground state. Such an evolution can be described by a differential equation
\begin{eqnarray}
d_{\tau}|\Psi(\tau)\rangle&=&-(\hat{H}-\langle\hat{H}\rangle)|\Psi(\tau)\rangle,
\label{eqn:ima2}
\end{eqnarray}
with the energy expectation $\langle\hat{H}\rangle=\langle\Psi(\tau)|\hat{H}|\Psi(\tau)\rangle$.
Thus, the imaginary-time evolution equation for the Gaussian state Eq.~(\ref{GS_state}) can be written as
\begin{eqnarray}
d_{\tau}|\Psi_{\rm GS}\rangle&=&-\hat{\mathcal{P}}(\hat{H}-E)|\Psi_{\rm GS}\rangle,\label{ima3}
\end{eqnarray}
where $E=\langle\Psi_{\rm GS}|\hat{H}|\Psi_{\rm GS}\rangle$ is the variational energy and $\hat{\mathcal{P}}$ is the projection operator onto the subspace spanned by tangent vectors of the variational manifold. The left-hand side of Eq.~(\ref{ima3}) gives
\begin{eqnarray}
d_{\tau}|\Psi_{\rm GS}\rangle = \hat{U}_{\rm GS}\hat{U}_{L}|0\rangle,
\label{eqn:left}
\end{eqnarray}
where the operator $\hat{U}_{L}$ is given by
\begin{eqnarray}
\hat{U}_{L}=\frac{1}{4}\text{:}\hat{\mathbf{A}}^{T}\mathbf{P}^{T}(\partial _{\tau}\mathbf{P})\hat{\mathbf{A}}\text{:}+\frac{i}{4}{\rm Tr}\left[\mathbf{P}^{T}(\partial _{\tau}\mathbf{P})\boldsymbol{\gamma}\right],
\end{eqnarray}
and $:\ :$ represents normal ordering with respect to the vacuum state. The right-hand side of Eq.~(\ref{ima3}) further reads
\begin{eqnarray}
-(\hat{H}-\langle\hat{H}\rangle)|\Psi_{\rm GS}\rangle=-\hat{U}_{\rm GS}\hat{U}_{R}|0\rangle,
\label{eqn:right}
\end{eqnarray}
where $\hat{U}_{R}=(i/4)\text{:}\hat{\mathbf{A}}^{T}\mathbf{P}^{T}\mathbf{h}\mathbf{P}\hat{\mathbf{A}}\text{:}+\delta\hat{H}$. Here, $\delta\hat{H}$ denotes the higher order terms of $c_{\mathbf{k}\sigma}$ that are orthogonal to the tangential
space which will be projected out by the $\hat{\mathcal{P}}$ operator in Eq.~(\ref{ima3}), and
\begin{eqnarray}
\mathbf{h}=4\frac{\delta E}{\delta\boldsymbol{\gamma}}
\label{hm}
\end{eqnarray}
is the functional derivative of the variational energy.
Comparing Eqs.~(\ref{eqn:left}) and (\ref{eqn:right}), and combining the covariant parameter defined by Eq.~(\ref{eqn:covariant}), we can finally obtain the imaginary time equation of motion (EOM) for the covariance matrix as
\begin{eqnarray}
\partial _{\tau}\boldsymbol{\gamma}=-\mathbf{h}-\boldsymbol{\gamma}\mathbf{h}\boldsymbol{\gamma}.
\label{eqn:EOM}
\end{eqnarray}

To evolve the variational parameter $\boldsymbol{\gamma}$ according to EOM given by Eq.~(\ref{eqn:EOM}), we need to calculate the functional derivative $\mathbf{h}$ defined in Eq.~(\ref{hm}). To accompalish that, we first rewrite the Hamiltonian Eq.(\ref{eqn:Hk}) with Majorana operators as
\begin{align}
\hat{H}=&\frac{1}{4}\sum_{\bf{k}\sigma}(\varepsilon_\mathbf{k}-\mu)
(2
+ ia^{\sigma}_{1,\mathbf{k}}a^{\sigma}_{2,\mathbf{k}} - ia^{\sigma}_{2,\mathbf{k}}a^{\sigma}_{1,\mathbf{k}}) \nonumber\\
&+\frac{U}{16V}\sum_{\bf{k,k',q}}(a_{1\bf{k}}^{\uparrow}-ia_{2\bf{k}}^{\uparrow})
(a_{1\bf{q-k}}^{\downarrow}-ia_{2\bf{q-k}}^{\downarrow}) \nonumber\\
& \quad \quad \quad \quad\times (a_{1\bf{q-k'}}^{\downarrow}+ia_{2\bf{q-k'}}^{\downarrow})
(a_{1\bf{k}'}^{\uparrow}+ia_{2\bf{k}'}^{\uparrow}).
\label{eqn:Ha}
\end{align}
Then using the formula of expectation under the fermionic Gaussian state as
\begin{eqnarray}
i^{p}\langle a_{j_1}\ldots a_{j_{2p}}\rangle_{\rm GS}=\rm{Pf}([\boldsymbol{\gamma}]_{j_1\ldots j_{2p}}),
\label{eqn:ExpectGauss}
\end{eqnarray}
where $1\leq j_1<\ldots<j_{2p}\leq4V$ and $\rm{Pf}([\boldsymbol{\gamma}]_{j_1\ldots j_{2p}})$ denotes the Pfaffian of the anti-symmetric matrix $[\boldsymbol{\gamma}]_{j_1\ldots j_{2p}}$ with rows and columns $j_1,\ldots,j_{2p}$ of $\boldsymbol{\gamma}$, we can calculate the variational energy $E=\langle\Psi_{\rm GS}|\hat{H}|\Psi_{\rm GS}\rangle$ and the functional derivative of the variational energy $\mathbf{h} = 4 \delta E/\delta\boldsymbol{\gamma}$. More details of the derivations are summarized in Appendix A and Appendix B.

In practical calculations, we can compute various physical observables from the covariant matrix $\boldsymbol{\gamma}$ of the ground state from the Gaussian variational method, which is determined by evolving Eqs.~(\ref{hm}) and (\ref{eqn:EOM}) until a convergence of variational energy $E$ and particle number $N_{\rm total}$ is reached. The latter can be expressed as
\begin{eqnarray}
N_{\text{total}}  = - {\frac{\partial E}{\partial\mu} }= V - \frac{1}{4}\sum_{i,j}\Omega_{i,j} \gamma_{i,j},
\label{eqn:num}
\end{eqnarray}
where $\Omega_{i,j}$ is the element of the symplectic matrix defined in Eq.~(\ref{fermi_sigma}). As a representative, we focus on the double occupancy defined as
\begin{align}
d &= \frac{1}{V} \sum_{\bf i} \langle c_{\bf i\uparrow}^\dag c_{\bf i\downarrow}^\dag c_{\bf i\downarrow} c_{\bf i\uparrow} \rangle_{\rm GS}.
\label{eqn:doubleoccupancy}
\end{align}
This quantity is of great interest in the context of superconductivity, since it describes the short-range pairing correlations of the ground state.
Besides, in dilute Fermi gases of ultracold atoms, double occupancy is found to be directly related to the contact parameter~\cite{Tan2008a,Tan2008b,Jensen2020,YuanYao2022}, which is a key quantity for the characterization of strongly interacting systems. Apparently, in strongly interacting limit, we have all particles with opposite spins paired and $d = \rho$, while in the weak-interaction limit, the double occupancy approaches its free fermion result $d = \rho^2$. Practically, we substitute the operators in Eq.~(\ref{eqn:doubleoccupancy}) by Majorana operators and combine with Eq.~(\ref{eqn:ExpectGauss}) to obtain the numerical result of $d$ using covariant matrix. The results are then compared with recent experiments of ultracold quantum gases in optical lattices obtained via site-resolved quantum gas microscopy~\cite{Mitra2018,Hartke2022}.

\begin{figure}[t]
\centering{}
\includegraphics[width=\linewidth]{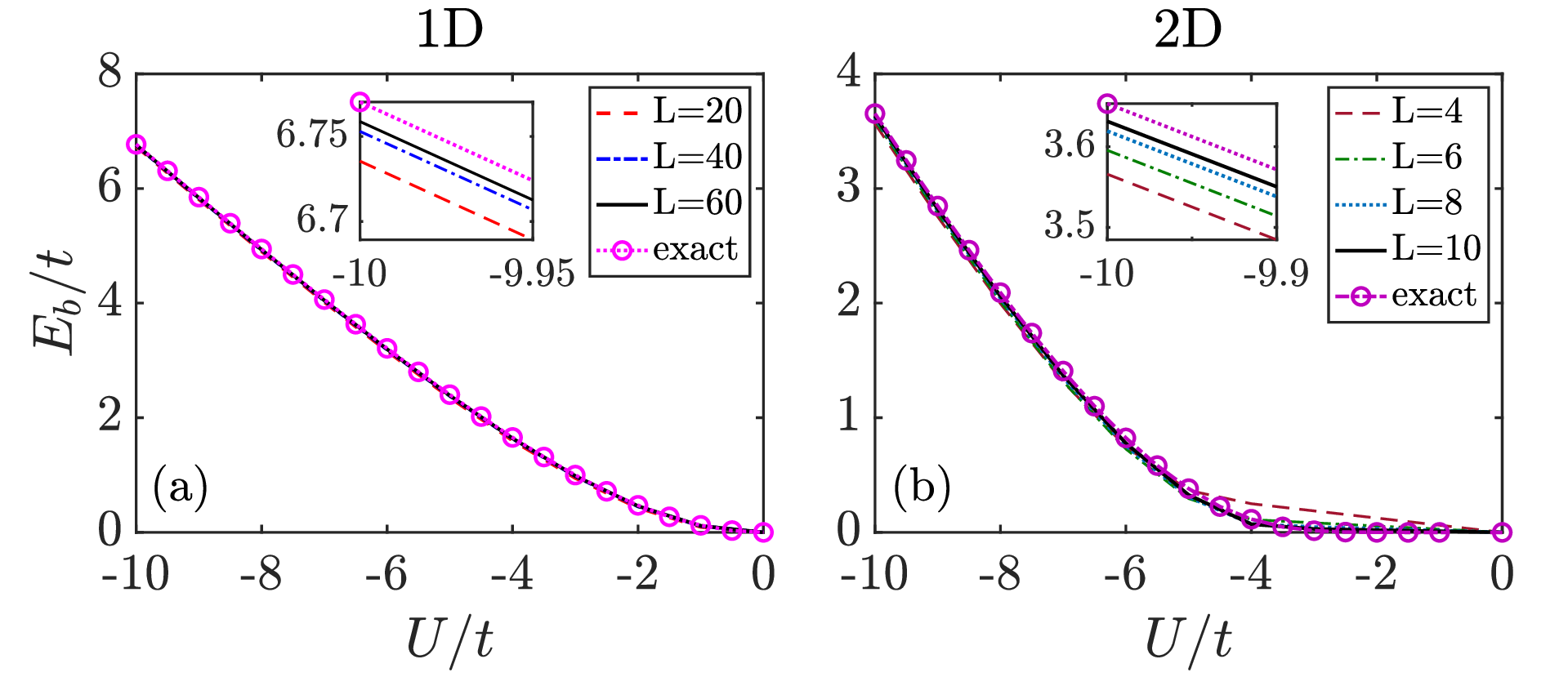}
\caption{(Color online) Two-body binding energy for (a) 1D chain of size $L=20,40,60$ and (b) 2D square lattice with linear system size $L=4,6,8,10$. Exact solutions for both cases are obtained by solving Eq.~(\ref{eqn:Eb}) numerically in the thermodynamic limit. The inset is a zoom in for strongly interacting limit, to show that the variational energy converges to the exact solution with increasing system size.}
\label{fig:Eb}
\end{figure}

\section{Ground state of Attractive Hubbard Model}
\label{sec:res}
\subsection{Two-body binding energy}
\label{sec:Eb}

We first consider a two-body problem and calculate the binding energy for different dimensions and lattice sizes, by tuning the chemical potential to fix the particle number as $N_\uparrow = N_\downarrow = 1$. This problem can be solved analytically, thus can provide a valuable benchmark for the dilute limit. The Hamiltonian of a two-body system can be written as
\begin{align}
\hat{H}^{(2)}=\sum_{\bf k,\sigma}\varepsilon_{\bf k}'c^\dagger_{\bf k\sigma}c_{\bf k\sigma} + \frac{U}{V} \sum_{\bf {k,k',q}}c^\dagger_{ {\bf q - k} \uparrow}c^\dagger_{{\bf k}\downarrow}c_{{\bf k'}\downarrow}c_{{\bf q-k'}\uparrow},
\label{eqn:EbH}
\end{align}
with the fermion number constraint $\sum_{\bf k} c_{\bf k \uparrow}^\dag c_{\bf k \uparrow} = \sum_{\bf k} c_{\bf k \downarrow}^\dag c_{\bf k \downarrow} = 1$. Here, a zero-point energy is introduced without loss of generality to shift the bottom of band to zero as $\varepsilon_{\bf k}' = 2D - 2\cos{\bf k}$. Since we are interested in the ground state, it is enough to consider a two-body wave function with zero center-of-mass momentum
\begin{align}
|\Psi^{(2)} \rangle &= \sum_{\bf k} \psi_{\bf k} c^\dagger_{\bf k \uparrow} c^\dagger_{ \bf -k \downarrow}  | 0\rangle.
\label{eqn:Psi2Body}
\end{align}
The Schr\"odinger equation then gives
\begin{align}
\hat{H}^{(2)} |\psi \rangle = E^{(2)} |\psi \rangle.
\label{eqn:Equ2Body}
\end{align}
Substituting Eq.~(\ref{eqn:Psi2Body}) into Eq.~(\ref{eqn:Equ2Body}), we can obtain the equation for coefficient $\psi_{\bf k}$ as
\begin{align}
2\varepsilon_{\bf k}' \psi_{\bf k} + \frac{U}{V} \sum_{\bf k'} \psi_{\bf k'} = E^{(2)} \psi_{\bf k}.
\end{align}
Finally, we arrive at the self-consistent equation for the two-body binding energy $E_b$ as
\begin{align}
-\frac{V}{U} = \sum_{\bf k}  \frac{1}{2 \varepsilon_{\bf k}' + E_b},
\label{eqn:Eb}
\end{align}
where $E_b$ is defined as $E_b = -E^{(2)}$.

\begin{figure}[t]
\centering{}
\includegraphics[width=\linewidth]{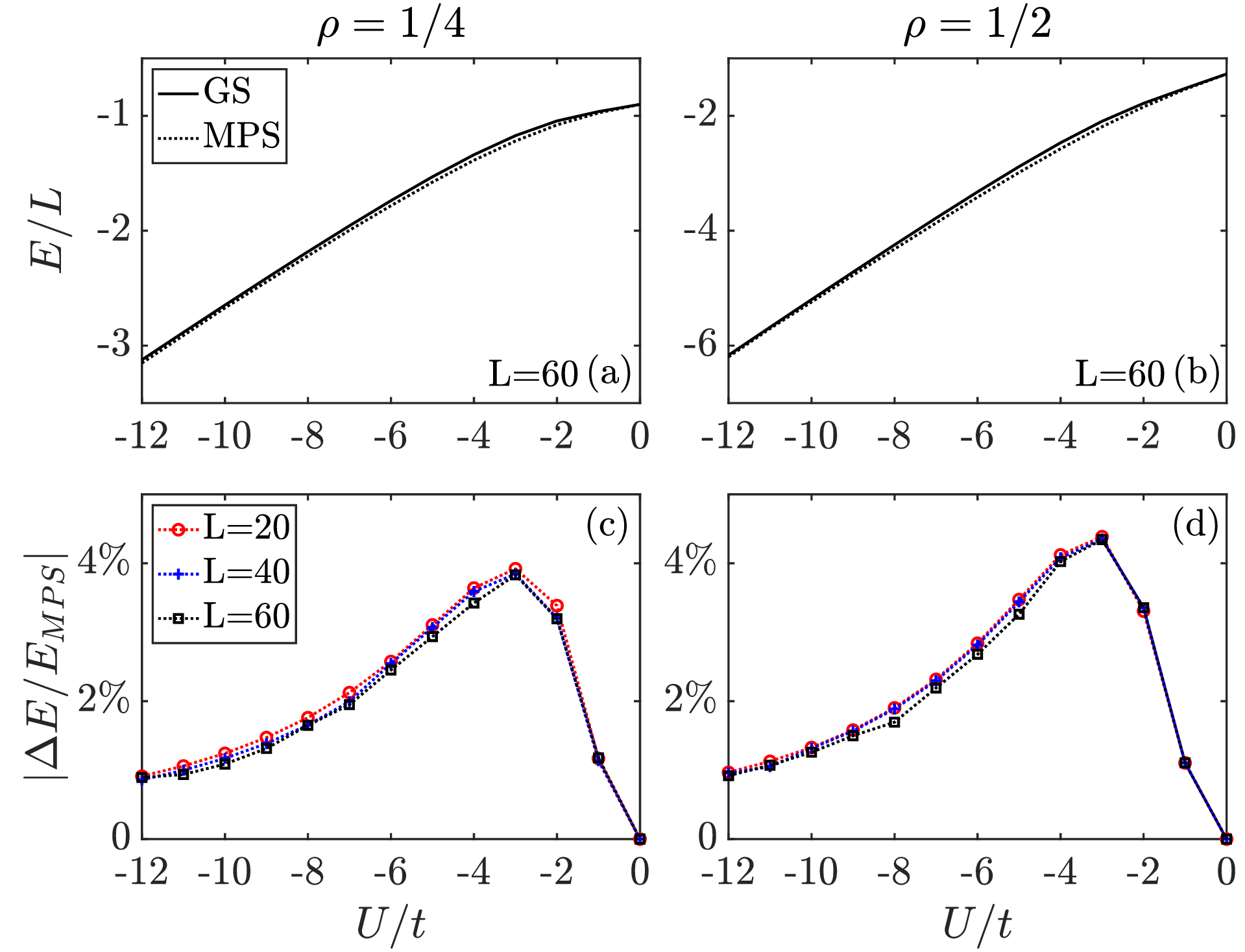}
\caption{(Color online) One-dimensional ground-state energy per-site for (a) $\rho = 1/4$ and (b) $\rho = 1/2$ and system size $L=60$, from the Gaussian variational method (solid) and MPS (dotted). Panel (c) and (d) show the systematic errors defined as $|(E_\text{MPS}-E_\text{GS})/E_\text{MPS}|$ for different size $L=20$ (red), $L=40$ (blue) and $L=60$ (black). }
\label{fig:1DEnergy}
\end{figure}

In Fig.~\ref{fig:Eb}, we show the two-body binding energy $E_b$ given by our Gaussian variational method (lines) for 1D and 2D systems, both using a BCS-like Gaussian state as the initial state for the imaginary-time evolution, and compare them with the corresponding exact results (dots). The numerical results agree well with the exact ones for both cases. By increasing system size $L$, the variational energy converges to the exact solution in the entire range of interaction, especially in the strongly interacting regime (insets of Fig.~\ref{fig:Eb}). These results demonstrate the validity of the Gaussian variational method in the dilute limit, and serve as the important guide for the following study.


\subsection{One-dimensional case}
\label{sec:1D}


We now consider a 1D chain of lengths $L= 20, 40, 60$ with periodic boundary condition and different fermion fillings $\rho = N/L = 1/4, 1/2$. This 1D system can be accurately solved by MPS~\cite{White1992,Michele2014}, which here serves as the benchmark for the Gaussian variational method. The maximum bond dimension for the MPS simulation is adopted as $M = 1600$, for which the energy variance $\langle\hat{H}^2 \rangle - \langle\hat{H}\rangle ^2$ is restricted within $10^{-6}$. A BCS-like Gaussian state is also used as the variational initial state.

In Figs.~\ref{fig:1DEnergy}(a) and \ref{fig:1DEnergy}(b), we present comparisons of the total energy per site versus the interaction strength from the Gaussian variational method and MPS for $\rho=1/4$ and $\rho=1/2$ with $L=60$, respectively. Correspondingly, the results of systematic error, defined as $|E_\text{MPS}-E_\text{GS}|/|E_\text{MPS}|$, for these two fillings with several different system sizes $L=20,40,60$ are shown in Figs.~\ref{fig:1DEnergy}(c) and \ref{fig:1DEnergy}(d). We find the results from our method are more accurate in both the weakly and strongly interacting regions. Instead, a relatively large systematic error ($\sim 4\%$) is observed in the intermediate interaction region. It is also interesting to find that the systematic error has little finite-size dependence as the results from different sizes almost collapse.



%
\begin{figure}[t]
\centering{}
\includegraphics[width=\linewidth]{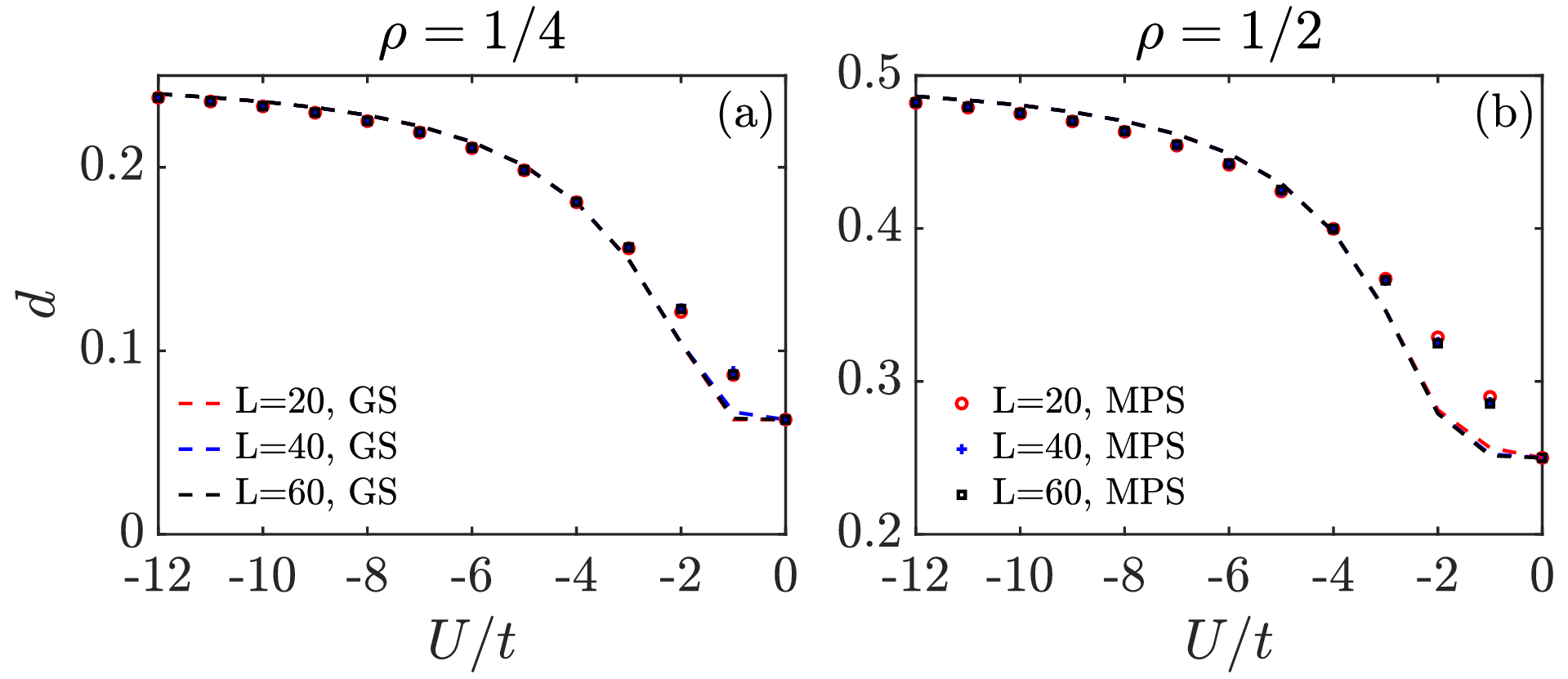}
\caption{(Color online) Double occupancy $d$ for a 1D chain with filling factor (a) $\rho = 1/4$ and (b) $\rho = 1/2$ from the Gaussian variational method (dashed lines), and MPS (symbols). The results share the same color codes with Fig.~\ref{fig:1DEnergy} as $L=20$ (red), $L=40$ (blue) and $L=60$ (black).}
\label{fig:1Ddoubleoccupancy}
\end{figure}

In Fig.~\ref{fig:1Ddoubleoccupancy}, we show comparisons of the ground-state double occupancy $d$ versus the interaction strength from the two methods. The results show good agreement in general, especially in the strongly interacting region, and again there is little finite-size dependence. The larger deviation in the weak interacting limit may be attributed to the fact that the ground state is better to be described by number-conserved states (shell effect). This may be improved in the future by combining Gaussian states with constraint of number conservation~\cite{Kraus2009}.


\subsection{Two-dimensional case}
\label{sec:2D}

In this section, we turn to the physically more relevant case of 2D system with different fermion fillings $\rho = N/L^2 = 1/4$ and $1/2$, and linear system sizes $L=4, 6, 8, 10$. For this case, the attractive Hubbard model can be solved in a numerically exact manner by QMC method, which can be used to benchmark our results.

\begin{figure}[t]
\centering{}
\includegraphics[width=\linewidth]{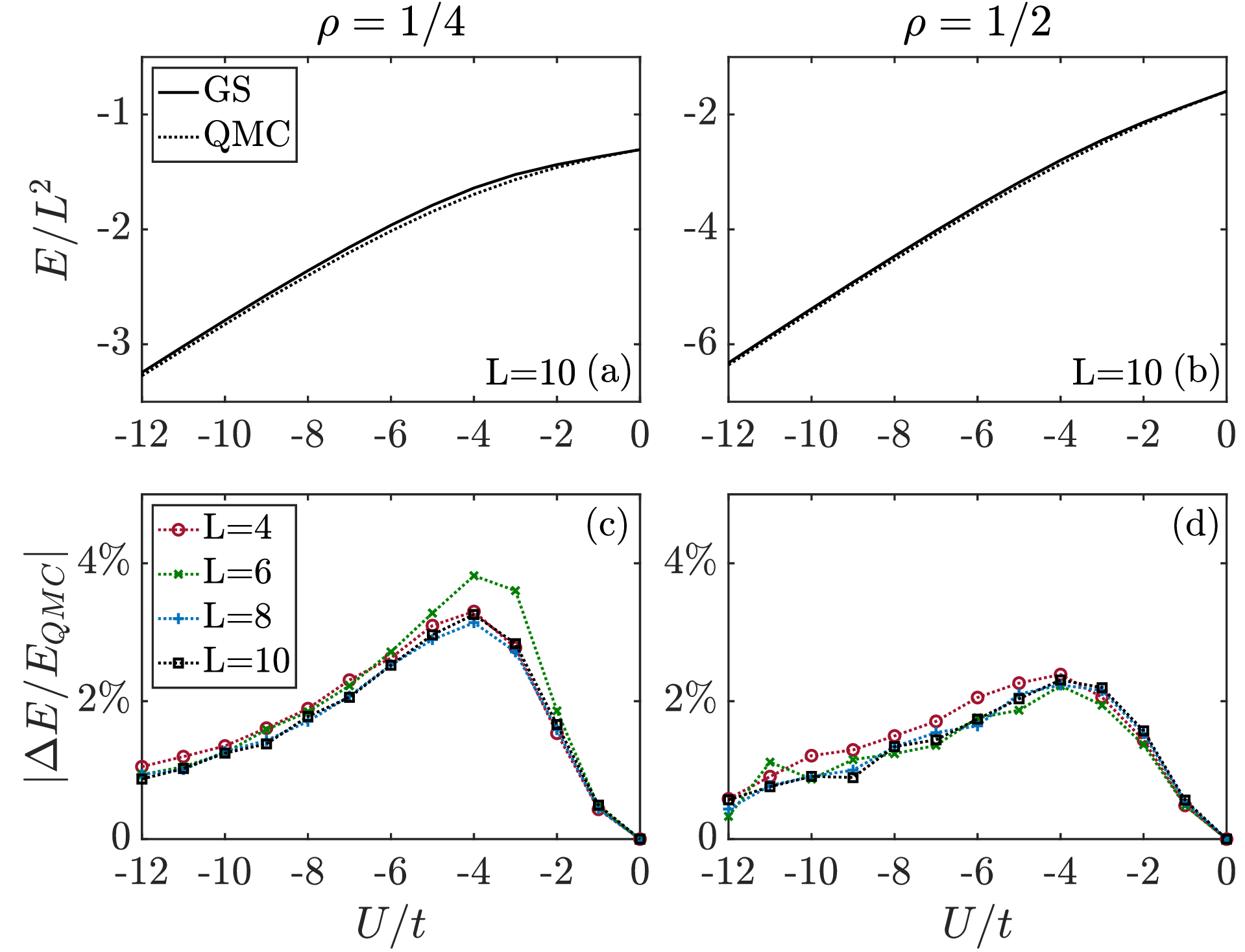}
\caption{(Color online) Two-dimensional ground-state energy per-site for a 2D square lattice with filling (a) $\rho = 1/4$ and (b) $\rho = 1/2$ and linear system size $L=10$, obtained from the Gaussian variational method (solid line) and QMC method (dashed line). Panels (c) and (d) show the systematic error defined as $|(E_\text{QMC}-E_\text{GS})/E_\text{QMC}|$ for different sizes $L=4$ (red), $L=6$ (green), $L=8$ (blue), and $L=10$ (black).}
\label{fig:2DEnergy}
\end{figure}

In Figs.~\ref{fig:2DEnergy}(a) and \ref{fig:2DEnergy}(b), we fix the system size as $L=10$ and vary the interaction $U$ to show the ground-state energy per site for $\rho=1/4$ and $1/2$, respectively. Good agreement is witnessed in both fillings throughout the entire interaction regime. In Figs~\ref{fig:2DEnergy}(c) and \ref{fig:2DEnergy}(d), we depict the systematic errors defined as $|E_\text{QMC}-E_\text{GS}|/|E_\text{QMC}|$ for different lattice sizes. Similar to the results presented in Fig.~\ref{fig:1DEnergy}, the systematic error reaches its maximum in the intermediate interaction regime with $U/t \sim -4$, with a largest value even smaller than $4\%$. This qualitative behavior is consistent with the observations in 2D continuum models, where an intermediate interaction strength falls into the strongly interacting region with the two-body binding energy comparable to the Fermi energy. For $\rho = 1/4$, the anomalous dependence on size of $L=6$ [green line in Fig.~\ref{fig:2DEnergy}(c)] comes from the closed-shell configuration~\cite{Bormann1991}.

\begin{figure}[t]
\centering{}
\includegraphics[width=\linewidth]{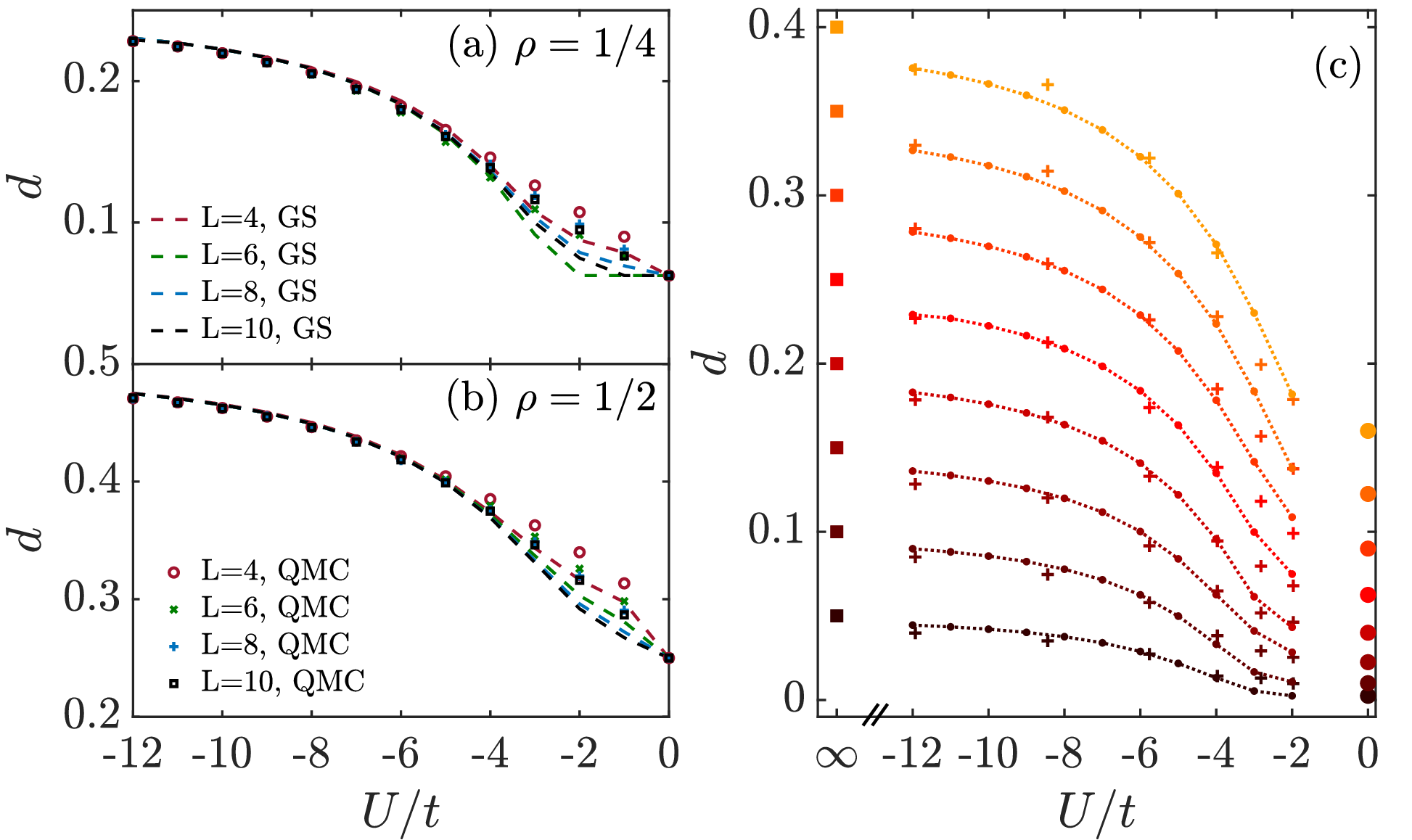}
\caption{(Color online) Double occupancy $d$ for a 2D square lattice with filling (a) $\rho = 1/4$ and (b) $\rho = 1/2$, obtained from the Gaussian variational method (dashed lines) and QMC (symbols). The results are presented with the same color code as in Fig.~ \ref{fig:2DEnergy} for $L=4$ (red), $L=6$ (green), $L=8$ (blue), and $L=10$ (black). (c) Double occupancy $d$ for a 2D attractive Hubbard model of linear size $L=10$ and particle filling  $\rho = 0.05 \sim 0.4$  with an interval of 0.05 (from bottom to top). The crosses, solid rectangles, and solid circles represent experiment data given by Ref.~\cite{Hartke2022} with $L=20$, and the small dots connected by dotted lines are numerical results. }
\label{fig:2Ddoubleoccupancy}
\end{figure}

We then calculate the double occupancy, and compare it with QMC results and experimental measurements. As can be seen in Figs.~\ref{fig:2Ddoubleoccupancy}(a) and \ref{fig:2Ddoubleoccupancy}(b), the double occupancy in 2D systems presents the same tendency as 1D chains, and shows size-dependence in weak and intermediate interaction regimes. For both 1D and 2D cases, our results show that the Gaussian variational method can well capture the short-range correlations quite well.
In Fig.~\ref{fig:2Ddoubleoccupancy}(c) we show the double occupancy for different fermion filling $\rho$ in $L=10$ system. The results from our Gaussian variational method are compared with the experimental data obtained for a cold gas with an effective attractive interaction in a 2D optical lattice~\cite{Hartke2022}. Although the experiment is performed at finite temperatures, good agreement is observed for all filling factors in the strongly interacting regime of $|U| > 4$, suggesting that the thermal fluctuations in experiments are negligible.

\section{Ground state of repulsive Hubbard model}
\label{sec:repul}
The repulsive Hubbard model can be studied by performing the particle-hole transformation to be mapped to an attractive model with spin polarization. To this aim, we first relax the equal spin population restriction in Eq.~(\ref{eqn:H}), and allow the spin-up and spin-down components to have different fillings $\rho_\uparrow$ and $\rho_\downarrow$. Then we introduce the partial particle-hole transformation of spin-down fermions, defined as $c_{\mathbf{i}\downarrow}^+\to(-1)^ic_{\mathbf{i}\downarrow}$ and $c_{\mathbf{i}\downarrow}\to(-1)^i c_{\mathbf{i}\downarrow}^+$. Under this transformation, we obtain a repulsive Hubbard model as

\begin{align}
\tilde{H}= &-t \sum_{\langle\mathbf{ij}\rangle, \sigma} \tilde{c}_{\mathbf{i} \sigma}^{\dagger} \tilde{c}_{\mathbf{j} \sigma}+\tilde{U} \sum_{\mathbf{i}} \tilde{c}_{\bf i \uparrow}^{\dagger} \tilde{c}_{\bf i \uparrow} \tilde{c}_{\bf i \downarrow}^{\dagger} \tilde{c}_{\bf i \downarrow},
\end{align}
where $\tilde{U}=-U>0$, and the particle filling $\tilde{\rho}_\uparrow = \frac{1}{V}\sum_{\mathbf{i}} \tilde{c}_{\mathbf{i} \uparrow}^{\dagger} \tilde{c}_{\mathbf{i} \uparrow} = \rho_\uparrow$ and $\tilde{\rho}_\downarrow = \frac{1}{V}\sum_{\mathbf{i}, } \tilde{c}_{\mathbf{i} \downarrow}^{\dagger} \tilde{c}_{\mathbf{i} \downarrow} = 1- \rho_\downarrow$. Here, we have omitted the trivial constant term in the transformed Hamiltonian. If the particle filling of the original model satisfies $\rho_\downarrow = 1-\rho_\uparrow$, we can obtain a repulsive Hubbard model with $\tilde{\rho}_\uparrow = \tilde{\rho}_\downarrow$.

\begin{figure}[t]
\centering{}
\includegraphics[width=\linewidth]{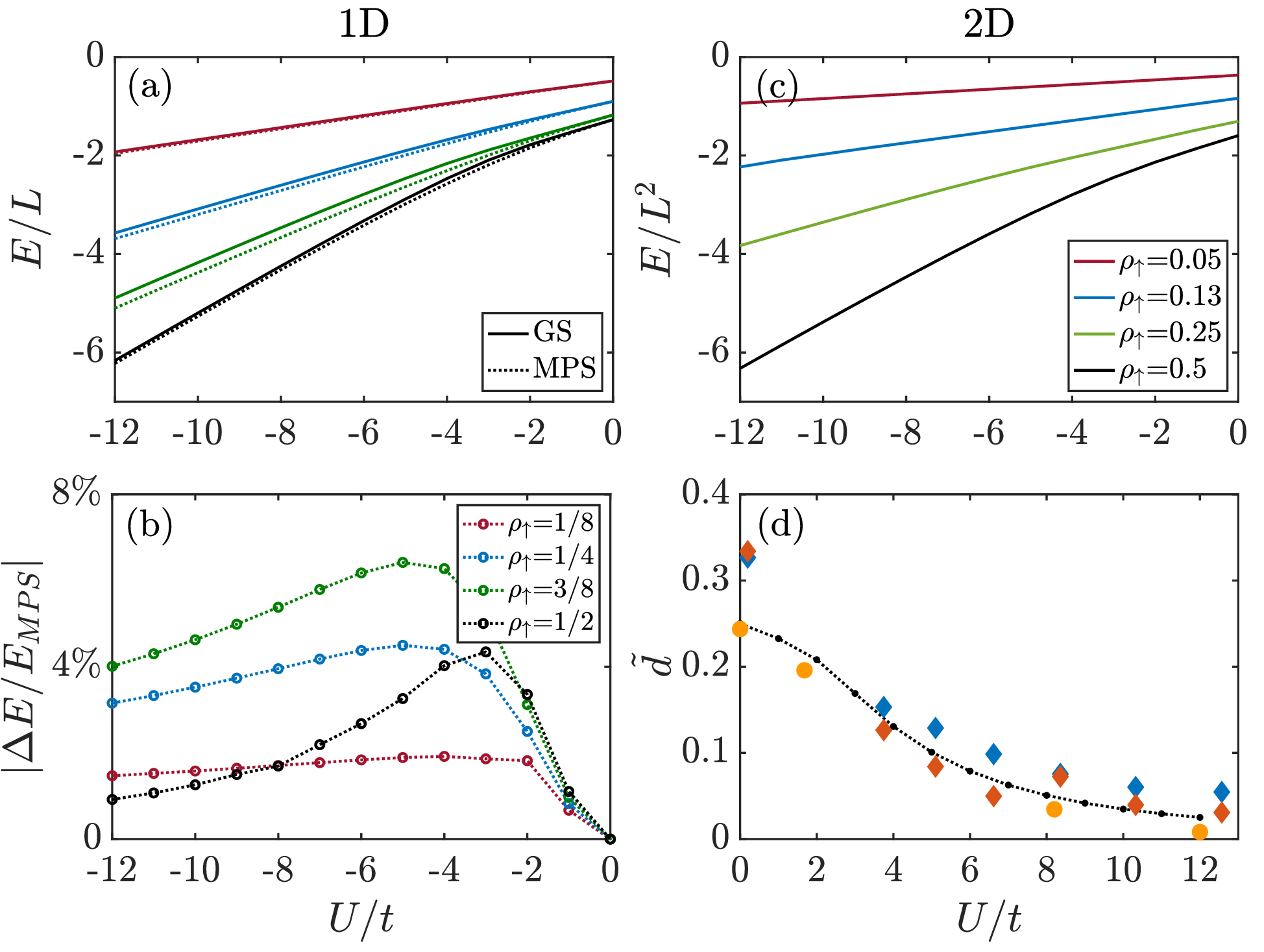}
\caption{(Color online) (a) Energy per-site of 1D attractive Hubbard model with different particle density $\rho_\uparrow = 1/8 \text{(red)}, 1/4\text{(blue)}, 3/8\text{(green)}, 1/2\text{(black)}$ and $\rho_\downarrow = 1-\rho_\uparrow$. Panel (b) gives the systematic errors compared to MPS results $|(E_\text{MPS}-E_\text{GS})/E_\text{MPS}|$. The system size is $L=40$ for $\rho_\uparrow = 1/8, 3/8$ and $L=60$ for   $\rho_\uparrow = 1/4, 1/2$.
(c) Energy per-site for 2D system with $L=10$, particle density $\rho_\uparrow= 0.05, 0.13, 0.25, 0.5$ and $\rho_\downarrow = 1-\rho_\uparrow$.
(d) Double occupancy $\tilde{d}$ for a 2D repulsive Hubbard model with $\rho = 1/2$ and $L=10$ (dotted lines), compared with experiments of Ref.~\cite{Boll2016} (diamonds) and Ref.~\cite{Kohl2016} (dots), respectively. The color of diamonds correspond to loose (blue) and tight (red) filter cases.
}
\label{fig:imbalance}
\end{figure}

We first consider a 1D system with four different sets of particle filling as $\rho_\uparrow = 1/8, 1/4, 3/8, 1/2$ and $\rho_\downarrow = 1-\rho_\uparrow$. In Fig.~\ref{fig:imbalance}(a) we compare our energy results (solid lines) with MPS results (dotted lines). The energy result shows ignorable size dependence, thus we use two different size $L=40$ and $L=60$ to fulfill the close shell configuration. We find the BCS initial state is no longer suitable for the calculation of imbalance systems, since it naturally assume the particle number of the two components to be equal. Here we use a random state as the initial state, the systematic errors displayed in Fig.~\ref{fig:imbalance}(b) show that the Gaussian-state approach behaves better in the highly imbalanced limit. 
 One of the possible reason for this large deviation is when the particle numbers are different, we need to consider the pairing with finite center-of-mass momentum, i.e. the FFLO state.  Also the systematic errors for both balance and imbalance case decrease while interactions are strong enough.
In Fig.~\ref{fig:imbalance}(c) we display the energy for a 2D system with $L=10$, and different particle density $\rho_\uparrow = 0.05, 0.13, 0.25, 0.50$ and $\rho_\downarrow = 1-\rho_\uparrow$. The Gaussian variational method is capable of calculating the spin-imbalanced system for both 1D and 2D case, as long as the particle number we consider fill up a degenerate energy level.
The energy for both 1D and 2D systems can be easily transformed as the energy of repulsive Hubbard model. For example, the energy of a repulsive Hubbard model with $\tilde{N}_\uparrow = \tilde{N}_\downarrow = V/4$ can be calculated as $\tilde{E} = E - U N_\uparrow$, where $E$ is the energy of an attractive Hubbard model with $N_\uparrow=V/4$ and $N_\downarrow=3V/4$.

We can also obtain the double occupancy of repulsive Hubbard model, which can be related to the double occupancy $d$ of the corresponding attractive model as
\begin{align}
\tilde{d} &=  \frac{1}{V} \sum_{\mathbf{i}} \langle \tilde{\Psi} | \tilde{c}_{\bf i \uparrow}^{\dagger} \tilde{c}_{\bf i \uparrow} \tilde{c}_{\bf i \downarrow}^{\dagger} \tilde{c}_{\bf i \downarrow} | \tilde{\Psi} \rangle
\nonumber\\
&= \rho_\uparrow -   \frac{1}{V} \sum_{\mathbf{i}} \langle c_{\bf i \uparrow}^{\dagger} c_{\bf i \uparrow}  c_{\bf i \downarrow}^\dagger c_{\bf i \downarrow} \rangle_{\rm GS}.
\end{align}
The double occupancy of attractive case can be transformed to a repulsive model. With that, we can compare our numerical calculation of half filling system with the experimental data obtained for a repulsive Hubbard model realized in cold quantum gases data adapted from Refs.~\cite{Boll2016}(diamonds) and \cite{Kohl2016}(dots), and demonstrate the results in Figs.~\ref{fig:imbalance}(d). We observe good agreement between numerical and most of the experimental data, except two points with very weak interaction in Refs.~\cite{Boll2016}, where the effective hole doping were introduced in the experiment.
	

\section{Conclusion and outlook}
\label{sec:con}

We have applied the Gaussian variational method to investigate the ground-state properties of 1D and 2D attractive Fermi-Hubbard model. We concentrate on the ground-state energy and double occupancy, which can be computed from the covariant matrix of the converged state after imaginary-time evolution. This method provides total energy with excellent agreement to numerically exact many-body methods, i.e., MPS in 1D and QMC in 2D, especially in strongly interacting region. The systematic errors have the same qualitative behavior for finite-size calculations, with a maximum in the intermediate region, and become smaller in both weak and strong interacting limits. In addition, the results of double occupancy given by the Gaussian variational method also show good agreement with those obtained by MPS and QMC methods. We then extend our calculation to repulsive Hubbard model, which can be studied by performing the particle-hole transformation to a spin polarized attractive Hubbard model. The Gaussian variation method is capable of calculating the ground state properties for an attractive Hubbard model with particle numbers fill up a degenerate energy level and satisfies $\rho_\downarrow = 1-\rho_\uparrow$. The double occupancy for a half filling repulsive Hubbard model also agrees well with experimental results.

Our results not only show that the Gaussian variational method can describe the attractive and repulsive Hubbard model with high accuracy, but also show that the Gaussian variational method has great potential for studying more complex systems. For example, this method can be straightforwardly generalized to spin- and mass-imbalanced Hubbard models, as well as multi-band Hubbard models. By considering a highly spin imbalance limit, we can study the impurity effect in half-filled Hubbard models, such as the magnetic polaron\cite{Koepsell2019,Brown2017}.The mass imbalance or multi-band effect may give rise to few-body correlations beyond the dominated two-body correlations\cite{Iskin2022}, which may lead to an even richer phase diagram\cite{Yuchi2021,liu2023quartet}. Another interesting issue is the Fermi polaron problem with strongly interacting two-component backgrounds\cite{Hui2022}, which may be relevant to the exciton-electron mixtures in transition metal dichalcogenides (TMDs) materials \cite{Tomasz2021}. Our results show the ability of the Gaussian variational method for studying the ground states of strongly interacting many-body systems to a very high precision and much less numerical cost.

	
\acknowledgments
We thank Tao Shi for helpful discussion, and Thomas Hartke and Martin Zwierlein for kindly sharing their experimental data. This work is supported by the National Natural Science Foundation of China (Grant No.~12074428, No.~92265208, No.~12047502), the Beijing Natural Science Foundation (Grant No.~Z180013), and the National Key R\&D Program of China (Grant No.~2018YFA0306501).
	
\appendix
\begin{widetext}
\section{Variational energy}
\label{Eexpression}

We use the expression given in Eq.~(\ref{eqn:Ha}) to determine the variational energy  $E=\langle\Psi_{\rm GS}|H|\Psi_{\rm GS}\rangle$, leading to
\begin{align}
E=&\langle\frac{1}{4}\sum_{\bf{k}\sigma}(\varepsilon_\mathbf{k} - \mu)
(2+ia^{\sigma}_
{1,\mathbf{k}}a^{\sigma}_{2,\mathbf{k}}-ia^{\sigma}_{2,\mathbf{k}}a^{\sigma}
_{1,\mathbf{k}}) \nonumber\\
&+\frac{U}{16V}\sum_{\bf{k,k',q}}(a_{1\bf{k}}^{\uparrow}-ia_{2\bf{k}}^{\uparrow})
(a_{1\bf{q-k}}^{\downarrow}-ia_{2\bf{q-k}}^{\downarrow})
(a_{1\bf{q-k'}}^{\downarrow}+ia_{2\bf{q-k'}}^{\downarrow})
(a_{1\bf{k}'}^{\uparrow}+ia_{2\bf{k}'}^{\uparrow})\rangle_{GS}\nonumber\\
=&\frac{1}{4}\sum_{\bf{k}\sigma}(\varepsilon_{\bf k}-\mu_\sigma)(2 + \gamma_{1,{\bf k}\sigma;2,{\bf k}\sigma}
-\gamma_{2,{\bf k}\sigma;1,{\bf k}\sigma}) + \frac{VU}{4}
+\frac{U}{8} \sum_{\bf{k}} ( \gamma_{1,{\bf k},\uparrow; 2,{\bf k},\uparrow} +   \gamma_{1,{\bf k},\downarrow; 2,{\bf k},\downarrow} - \gamma_{2,{\bf k},\downarrow; 1,{\bf k},\downarrow} - \gamma_{2,{\bf k},\uparrow; 1,{\bf k},\uparrow}) \nonumber\\
&+\frac{U}{16V} \sum_{\bf{k, k', q}} \bigg[
(\gamma_{1,{\bf k},\uparrow; 1,{\bf q-k},\downarrow} - \gamma_{2,{\bf k},\uparrow; 2,{\bf q-k},\downarrow})(\gamma_{2,{\bf q-k'},\downarrow; 2,{\bf k'},\uparrow} - \gamma_{1,{\bf q-k'},\downarrow; 1,{\bf k'},\uparrow}) \nonumber\\
&\quad \quad \quad \quad \quad \quad + (\gamma_{1,{\bf k},\uparrow; 1,{\bf q-k'},\downarrow} + \gamma_{2,{\bf k},\uparrow; 2,{\bf q-k'},\downarrow})(\gamma_{1,{\bf q-k},\downarrow; 1,{\bf k'},\uparrow} + \gamma_{2,{\bf q-k},\downarrow; 2,{\bf k'},\uparrow})\nonumber\\
&\quad \quad \quad \quad \quad \quad -  (\gamma_{1,{\bf k},\uparrow; 1,{\bf k'},\uparrow} + \gamma_{2,{\bf k},\uparrow; 2,{\bf k'},\uparrow})(\gamma_{1,{\bf q-k},\downarrow; 1,{\bf q-k'},\downarrow} + \gamma_{2,{\bf q-k},\downarrow; 2,{\bf q-k'},\downarrow} )\nonumber\\
&\quad \quad \quad \quad \quad \quad +(\gamma_{1,{\bf k},\uparrow; 2,{\bf q-k'},\downarrow} - \gamma_{2,{\bf k},\uparrow; 1,{\bf q-k'},\downarrow})(\gamma_{2,{\bf q-k},\downarrow; 1,{\bf k'},\uparrow} - \gamma_{1,{\bf q-k},\downarrow; 2,{\bf k'},\uparrow}) \nonumber\\
&\quad \quad \quad \quad \quad \quad - (\gamma_{1,{\bf k},\uparrow; 2,{\bf q-k},\downarrow} + \gamma_{2,{\bf k},\uparrow; 1,{\bf q-k},\downarrow})(\gamma_{1,{\bf q-k'},\downarrow; 2,{\bf k'},\uparrow} + \gamma_{2,{\bf q-k'},\downarrow; 1,{\bf k'},\uparrow})\nonumber\\
&\quad \quad \quad \quad \quad \quad +  (\gamma_{1,{\bf k},\uparrow; 2,{\bf k'},\uparrow} - \gamma_{2,{\bf k},\uparrow; 1,{\bf k'},\uparrow})(\gamma_{1,{\bf q-k},\downarrow; 2,{\bf q-k'},\downarrow} - \gamma_{2,{\bf q-k},\downarrow; 1,{\bf q-k'},\downarrow} ) \bigg]
\label{E}.
\end{align}
The energy we show in the main text is obtained via this expression.

\section{Derivative of the variational energy}
\label{partialEexpression}

After we get Eq.~(\ref{E}), we can calculate the functional derivative $h = 4 \delta E/ \delta \gamma$. The element of matrix $h$ can be expressed as $h_{s_1,{\bf k_1}, \sigma_1; s_2, {\bf k_2},\sigma_2} = 4 \delta E/ \delta\gamma_{s_1,{\cp k_1}, \sigma_1; s_2,{\cp k_2},\sigma_2}$, where $s_1$ and $s_2$ are the indices of Majorana operators (1 or 2). The expression is given by
\begin{align}
&h_{s_1,{\bf k_1}\sigma_1; s_2,{\bf k_2}\sigma_2}=4\frac{\delta E}{\delta\gamma_{s_1,{\cp k_1},\sigma_1; s_2,{\cp k_2},\sigma_2}}\nonumber\\
&= - \delta_{s_1,1} \delta_{s_2,1}\delta_{\sigma_1, \uparrow} \delta_{\sigma_2, \uparrow} \sum_{\cp k}(\gamma_{1, {\bf k}, \downarrow; 1,{\bf k+k_1-k_2},\downarrow} + \gamma_{2, {\bf k}, \downarrow; 2,{\bf  k+k_1-k_2},\downarrow}) \nonumber\\
&\quad + \delta_{s_1,1} \delta_{s_2,1}\delta_{\sigma_1, \uparrow} \delta_{\sigma_2, \downarrow} \sum_{\cp k}(\gamma_{2, {\bf k_1+k_2-k}, \downarrow; 2,\cp k,\uparrow} - \gamma_{1, {\bf k_1+k_2-k}, \downarrow; 1,\cp k,\uparrow}
+ \gamma_{1, {\bf k}, \downarrow; 1, {\bf k+k_1-k_2},\uparrow} + \gamma_{2, {\bf k}, \downarrow; 2,{\bf k+k_1-k_2},\uparrow} )\nonumber\\
&\quad + \delta_{s_1,1} \delta_{s_2,1}\delta_{\sigma_1, \downarrow} \delta_{\sigma_2, \uparrow} \sum_{\cp k}(-\gamma_{1, {\bf k}, \uparrow; 1, {\bf k_1+k_2-k},\downarrow} + \gamma_{2, {\bf k}, \uparrow; 2,{\bf k_1+k_2-k},\downarrow} + \gamma_{1, {\bf k}, \uparrow; 1,{\bf k+k_1-k_2},\downarrow} + \gamma_{2, {\bf k}, \uparrow; 2,{\bf  k+k_1-k_2},\downarrow} )\nonumber\\
&\quad - \delta_{s_1,1} \delta_{s_2,1}\delta_{\sigma_1, \downarrow} \delta_{\sigma_2, \downarrow}\sum_{\cp k}(\gamma_{1, {\bf k}, \uparrow; 1,{\bf k+k_1-k_2},\uparrow} + \gamma_{2, {\bf k}, \uparrow; 2,{\bf k+k_1-k_2},\uparrow})\nonumber\\
&\quad + \delta_{s_1,1} \delta_{s_2,2}\delta_{\sigma_1, \uparrow} \delta_{\sigma_2, \uparrow}\bigg[\sum_{\cp k}\delta_{\bf k_1,k_2} ( \varepsilon_{\bf k} - \mu +  \frac{U}{2}  )+\sum_{\cp k}(\gamma_{1, {\bf k}, \downarrow; 2,{\bf k+k_1-k_2},\downarrow} - \gamma_{2, {\bf k}, \downarrow; 1,{\bf k+k_1-k_2},\downarrow}) \bigg] \nonumber\\
&\quad +  \delta_{s_1,1} \delta_{s_2,2}\delta_{\sigma_1, \uparrow} \delta_{\sigma_2, \downarrow}\sum_{\cp k}(\gamma_{2, {\bf k}, \downarrow; 1,{\bf k+k_1-k_2},\uparrow} - \gamma_{1, {\bf k}, \downarrow; 2,{\bf k+k_1-k_2},\uparrow} -\gamma_{1, {\bf k_1+k_2-k}, \downarrow; 2,{\bf k},\uparrow} - \gamma_{2, {\bf k_1+k_2-k}, \downarrow; 1,{\bf k},\uparrow}) \nonumber\\
&\quad + \delta_{s_1,1} \delta_{s_2,2}\delta_{\sigma_1, \downarrow} \delta_{\sigma_2, \uparrow} \sum_{\cp k}(\gamma_{2, {\bf k}, \uparrow; 1,{\bf k+k_1-k_2},\downarrow} - \gamma_{1, {\bf k}, \uparrow; 2,{\bf k+k_1-k_2},\downarrow} - \gamma_{1, {\bf k}, \uparrow; 2,{\bf k_1+k_2-k},\downarrow} - \gamma_{2, {\bf k}, \uparrow; 1,{\bf k_1+k_2-k},\downarrow})  \nonumber\\
&\quad + \delta_{s_1,1} \delta_{s_2,2}\delta_{\sigma_1, \downarrow} \delta_{\sigma_2, \downarrow}\bigg[\sum_{\cp k} \delta_{\bf k_1,k_2} ( \varepsilon_{\bf k} - \mu +  \frac{U}{2}  ) + \sum_{\cp k}(\gamma_{1, {\bf k}, \uparrow; 2,{\bf k+k_1-k_2},\uparrow} - \gamma_{2, {\bf k}, \uparrow; 1,{\bf k+k_1-k_2},\uparrow}) \bigg] \nonumber\\
&\quad - \delta_{s_1,2} \delta_{s_2,1}\delta_{\sigma_1, \uparrow} \delta_{\sigma_2, \uparrow}\bigg[\sum_{\cp k} \delta_{\bf k_1,k_2} ( \varepsilon_{\bf k} - \mu +  \frac{U}{2}  ) + \sum_{\cp k}(\gamma_{1, {\bf k}, \downarrow; 2,{\bf k+k_1-k_2},\downarrow} - \gamma_{2, {\bf k}, \downarrow; 1,{\bf k+k_1-k_2},\downarrow}) \bigg] \nonumber\\
&\quad + \delta_{s_1,2} \delta_{s_2,1}\delta_{\sigma_1, \uparrow} \delta_{\sigma_2, \downarrow}  \sum_{\cp k}(\gamma_{1, {\bf k}, \downarrow; 2,{\bf k+k_1-k_2},\uparrow} - \gamma_{2, {\bf k}, \downarrow; 1,{\bf k+k_1-k_2},\uparrow} - \gamma_{1, {\bf k_1+k_2-k}, \downarrow; 2,{\bf k},\uparrow} - \gamma_{2, {\bf k_1+k_2-k}, \downarrow; 1,{\bf k},\uparrow}) \nonumber\\
&\quad + \delta_{s_1,2} \delta_{s_2,1}\delta_{\sigma_1, \downarrow} \delta_{\sigma_2, \uparrow} \sum_{\cp k}(\gamma_{1, {\bf k}, \uparrow; 2,{\bf k+k_1-k_2},\downarrow} - \gamma_{2, {\bf k}, \uparrow; 1,{\bf k+k_1-k_2},\downarrow} - \gamma_{1, {\bf k}, \uparrow; 2,{\bf k_1+k_2-k},\downarrow} - \gamma_{2, {\bf k}, \uparrow; 1,{\bf k_1+k_2-k},\downarrow}) \nonumber\\
&\quad - \delta_{s_1,2} \delta_{s_2,1}\delta_{\sigma_1, \downarrow} \delta_{\sigma_2, \downarrow} \bigg[ \sum_{\cp k} \delta_{\bf k_1,k_2} ( \varepsilon_{\bf k} - \mu +  \frac{U}{2}  ) + \sum_{\cp k}(\gamma_{1, {\bf k}, \uparrow; 2,{\bf k+k_1-k_2},\uparrow} - \gamma_{2, {\bf k}, \uparrow; 1,{\bf k+k_1-k_2},\uparrow}) \bigg] \nonumber\\
&\quad - \delta_{s_1,2} \delta_{s_2,2}\delta_{\sigma_1, \uparrow} \delta_{\sigma_2, \uparrow}\sum_{\cp k}(\gamma_{1, {\bf k}, \downarrow; 1,{\bf k+k_1-k_2},\downarrow} + \gamma_{2, {\bf k}, \downarrow; 2,{\bf  k+k_1-k_2},\downarrow})\nonumber\\
&\quad + \delta_{s_1,2} \delta_{s_2,2}\delta_{\sigma_1, \uparrow} \delta_{\sigma_2, \downarrow} \sum_{\cp k}( \gamma_{1, {\bf k_1+k_2-k}, \downarrow; 1,\cp k,\uparrow} - \gamma_{2, {\bf k_1+k_2-k}, \downarrow; 2,\cp k,\uparrow} + \gamma_{1, {\bf k}, \downarrow; 1, {\bf k+k_1-k_2},\uparrow} + \gamma_{2, {\bf k}, \downarrow; 2,{\bf k+k_1-k_2},\uparrow}) \nonumber\\
&\quad + \delta_{s_1,2} \delta_{s_2,2}\delta_{\sigma_1, \downarrow} \delta_{\sigma_2, \uparrow} \sum_{\cp k}(\gamma_{1, {\bf k}, \uparrow; 1, {\bf k_1+k_2-k},\downarrow} - \gamma_{2, {\bf k}, \uparrow; 2,{\bf k_1+k_2-k},\downarrow} + \gamma_{1, {\bf k}, \uparrow; 1,{\bf k+k_1-k_2},\downarrow} + \gamma_{2, {\bf k}, \uparrow; 2,{\bf  k+k_1-k_2},\downarrow}) \nonumber\\
&\quad - \delta_{s_1,2} \delta_{s_2,2}\delta_{\sigma_1, \downarrow} \delta_{\sigma_2, \downarrow} \sum_{\cp k}(\gamma_{1, {\bf k}, \uparrow; 1,{\bf k+k_1-k_2},\uparrow} + \gamma_{2, {\bf k}, \uparrow; 2,{\bf k+k_1-k_2},\uparrow}).
\end{align}
\end{widetext}

\bibliographystyle{ref}
\bibliography{Gaussian_Hubbard_arxiv}

\end{document}